\documentclass[reprint, superscriptaddress, amsmath,amssymb, aps, prresearch, longbibliograph]{revtex4-2}

\usepackage{amsmath}
\usepackage{graphicx}
\usepackage{ulem}
\usepackage{bm}
\usepackage{natbib}
\usepackage{dsfont}
\usepackage{epsfig}
\usepackage{feynmf}
\usepackage{blindtext, rotating}
\usepackage{mathtools}
\usepackage{dsfont}
\usepackage{subcaption}
\usepackage{physics}
\usepackage{amsfonts}
\usepackage{ragged2e}
\usepackage{siunitx}
\usepackage{comment}
\usepackage{soul}
\usepackage{lipsum}

\usepackage{xcolor} 
\newcommand{\normord}[1]{:\mathrel{#1}:}

\DeclareCaptionJustification{justified}{\justifying}

\begin{document}

\title{Nonlinearities in Black Hole Ringdowns and the Quantization of Gravity}

\author{Thiago Guerreiro}
\email{barbosa@puc-rio.br}
\affiliation{Department of Physics, Pontifical Catholic University of Rio de Janeiro, Rio de Janeiro 22451-900, Brazil}

\begin{abstract}
Einstein’s theory of gravity admits a low energy effective quantum field description from which predictions beyond classical general relativity can be drawn. As gravitational wave detectors improve, one may ask whether non-classical features of such theory can be experimentally verified.
Here we argue that nonlinear effects in black hole ringdowns can be sensitive to the graviton number statistics and other quantum properties of gravitational wave states. The prediction of ringdown signals, potentially measurable in the near future, might require the inclusion of quantum effects. This offers a new route to probing the quantum nature of gravity and gravitational wave entanglement.
\end{abstract}


\maketitle

\textit{Introduction.} --
Experimentally detecting a quantum gravitational effect is such an outstanding challenge that even the need for a quantum theory of gravity has been cast into doubt. 
Freeman Dyson, for instance, raised the question of whether gravitons are \textit{in principle} detectable \cite{dyson2014graviton}. Penrose suggested that gravity enforces quantum state reduction, which would explain how objective classical reality emerges out of coherent superpositions \cite{penrose1996gravity}. Others have pointed out connections between Einstein's theory and thermodynamics \cite{bekenstein1973black}, supporting the view that gravity and space-time itself are emergent phenomena analogous to sound waves \cite{jacobson1995thermodynamics, padmanabhan2010thermodynamical} or entropic forces \cite{verlinde2011origin}. 
The list goes on and on, with the main point being that any hint on the quantum nature of gravity, either in the weak or strong field regimes, is of incalculable value.
 

In the past couple of years, the interest in experiments aiming to establish the quantum nature of gravity has increased significantly \cite{aspelmeyer2022avoid}.
While directly detecting single gravitons might be impossible due to the smallness of gravitational scattering cross sections \cite{dyson2014graviton}, quantum gravitational effects could manifest in cosmological observations \cite{krauss2014}, tabletop experiments \cite{bose2017spin, marletto2017gravitationally, carney2022newton} or in the statistical properties of quantum gravitational wave states (GWs) described by low energy effective field theory \cite{ guerreiro2020gravity, parikh2021a}. In effect, quantum fluctuations of GWs induce noise which could be measured through the separation of test masses \cite{parikh2021a, parikh2021b} or by monitoring the optical state present in an interferometer \cite{guerreiro2020gravity}. 
These quantum fluctuations can be present in states with macroscopic mean number of gravitons and could in principle be measured using GW interferometers, which we may regard as a kind of homodyne detector \cite{guerreiro2022quantum}. 

Recently, nonlinear effects during black hole (BH) ringdown were spotted in numerical simulations \cite{mitman2023nonlinearities, cheung2023nonlinear}. In quantum optics, nonlinear field interactions generate non-classical states \cite{hillery1985conservation, albarelli2016nonlinearity}, hence one may ask whether GW nonlinearities could lead to quantum gravitational effects potentially measurable in GW detectors.
Here, we argue on physical grounds that nonlinearities in BH ringdowns can be sensitive to the graviton number statistics and other quantum properties of GW states, and hence provide a smoking gun to the quantum nature of gravity. Conversely, nonlinearities also provide mechanisms for the generation of non-classical GW states. 
Ringdowns will be accessible with next-generation GW detectors \cite{abbott2020prospects, maggiore2020science, evans2021horizon} and predicting the strength of these signals might require the inclusion of quantum effects. 
Phenomenological discrepancies between classical expectations and measurements of ringdowns could guide the development of effective quantum field theories of strong gravity.


In the following, we briefly review the recent numerical evidence for nonlinear effects in BH ringdowns and present a physical argument 
for the dependence of nonlinear ringdown signals on quantum properties of the parent GW states. We discuss how the graviton number and quadrature of nonlinear quantum GWs depend on amplitude-phase correlations of the parent mode and calculate corrections to the classical prediction for a class of quantum GW states, namely squeezed-coherent-thermal states.
Strong quantum nonlinear interactions generically lead to mode entanglement.
We then show how gravitational entanglement could in principle be witnessed using quantum-enhanced GW detectors. We conclude with a discussion on the prospects of generating non-classical GWs, particularly at the end stages of binary BH mergers.

\textit{Frequency doubling.} -- 
Consider quasinormal modes (QNMs) of a perturbed Kerr BH indexed by angular harmonic and overtone numbers $ (\ell,m,n) $. Through numerical simulations, the authors of \cite{mitman2023nonlinearities} show that for a typical ringdown following the merging of two BHs, over a wide range of initial mass ratios, a parent mode $(2,2,0)$ with amplitude $ A_{(2,2,0)} $ and frequency $ \omega_{(2,2,0)} $ gives birth to a second-order mode $ (4,4,0) $ with amplitude $ A_{(4,4)}^{(2,2,0)\times(2,2,0)} \propto (A_{(2,2,0)})^{2} $ and frequency $ \omega_{(2,2,0)\times(2,2,0)} = 2\omega_{(2,2,0)} $. The amplitude of the second-order $ (4,4,0) $ mode is comparable or even larger than the corresponding linear $(4,4)$ component and the proportionality constant between $ A_{(4,4)}^{(2,2,0)\times(2,2,0)} $ and $ (A_{(2,2,0)})^{2} $ is of order unity, suggesting strong coupling. Consistent results are found in \cite{cheung2023nonlinear}, where it is also shown that the phase of the nonlinear wave is twice that of its parent plus a constant shift.
Nonlinear effects, particularly frequency doubling, are ubiquitous in BH ringdown. 

Doubling of the frequency, phase and a quadratic dependence of the nonlinear mode on the parent amplitude are familiar to quantum optics. 
The effective field theory interaction Hamiltonian describing frequency doubling in a cavity is
\begin{equation}
    H_{ab} 
    = \kappa   aab^{\dagger} + \kappa^{*}  a^{\dagger}a^{\dagger} b  , 
    \label{hamiltonian}
\end{equation}
where $ \kappa \in \mathbb{C} $ is the coupling constant and $ a^{\dagger} (a), b^{\dagger} (b) $ are single particle creation (annihilation) operators of the parent and nonlinear modes, respectively \footnote{Note free evolution of the field and eventual quantum jump operators corresponding to decay also contribute to the dynamics.}. 
Particle number conservation -- valid for times much shorter than the inverse damping rate -- requires that the Hamiltonian describing frequency doubling commutes with
$ J = N_{a} + 2N_{b} $, where $ N_{a,b}$ denotes the number of particles in each mode. Using the canonical commutation relations we can show that the only third-order Hermitian operator that satisfies this condition is $ H_{ab}$ (see the Appendix). 
Including dissipation, the Hamiltonian \eqref{hamiltonian} leads to the parametric amplifier equations for frequency doubling \cite{mandel1982squeezing},
\begin{eqnarray}
    \dot{\alpha} &=& -2i\kappa \beta \alpha^{\dagger} - \gamma_{a}\alpha \label{OPA1} \\
    \dot{\beta} &=& -i\kappa \alpha^{2} - \gamma_{b}\beta \label{OPA2}
\end{eqnarray}
where $ \gamma_{a,b} $ are the damping rates, $ \alpha = ae^{i\omega t} $ and $ \beta = be^{2i\omega t} $.
In the following we assume $ \kappa $ is real for simplicity, but this can be extended to arbitrary complex values.

Black hole ``vibrations'' can be understood as GWs undergoing unstable orbits around the horizon \cite{goebel1972, yang2012quasinormal}. We can think of BHs as providing a temporary storage of GWs similarly to a cavity, where the real part of the QNM frequency corresponds to the resonance and the imaginary part accounts for graviton absorption. 
On physical grounds, we postulate that ringdown frequency doubling can be modeled as damped quantum harmonic oscillators \cite{maggiore2008physical} governed by the effective parametric amplifier equations \eqref{OPA1} and \eqref{OPA2}.

Note the relations between nonlinear interactions of Kerr perturbations and parametric oscillators have been previously discussed in \cite{yang2015turbulent}, 
and bosonic models for interacting GWs have been considered in \cite{sawyer2020quantum} for describing properties concerning particle number statistics \footnote{For a discussion on the canonical quantization of gravitational perturbations of a Kerr black hole see also \cite{candelas1981quantization}.}.

\textit{Dependence on graviton statistics.} -- It has long been recognized that the properties of a nonlinear mode are dependent on the particle number statistics of its parent state \cite{ekert1988second, kozierowski1977quantum}; see also \cite{olsen2002dynamical} and references therein for more general discussions on parametric processes beyond frequency doubling. The physical argument is simple and goes beyond quadratic field effects: particle number statistics is related to the bunching or anti-bunching character of the field \cite{davidovich1996sub}. Since a nonlinear interaction requires two or more particles, bunched states are converted more efficiently than anti-bunched states \footnote{I acknowledge A. Z. Khoury for pointing me this argument.}. 

In the case of frequency doubling, if mode $ b $ is initially in the vacuum, Eqs. \eqref{OPA1} and \eqref{OPA2} imply the mean number of particles $ \langle N_{b} \rangle $ scales with interaction time $ t $ as \cite{ekert1988second}
\begin{equation}
   \langle N_{b}(t) \rangle = g_{2}  \ \langle N_{a}(0) \rangle^{2} \ \kappa^{2}t^{2} , 
   \label{growth}
\end{equation}
where,
\begin{equation}
    g_{2} = \frac{\langle \normord{N_{a}^{2}(0)} \rangle}{ \langle N_{a}(0) \rangle^{2}} 
\end{equation}
is the second-order field auto-correlation function \footnote{$\normord{\mathcal{O}}$ denotes normal ordering.}.
Eq. \eqref{growth} is valid for $ \sqrt{2g_{2} \langle N_{a}(0) \rangle} \ \kappa t \ll 1 $ and $t \ll  \gamma^{-1}_{a}, \gamma^{-1}_{b} $.

In a quantum setting, the growth of frequency doubled GWs is sensitive to the graviton statistics of the initial parent state. 
Depending on the quantum state of the parent mode, the generated nonlinear signal will be more or less intense. For coherent states, the closest to classical waves, we have $ g_{2} = 1 $. For thermal states $ g_{2} = 2 $, while squeezed states have $ g_{2} = 3 + \frac{1}{\langle N_{a}(0)\rangle} $. 
Nonlinear GW production for parent squeezed and thermal states is significantly enhanced compared to coherent GWs.

Current detectors such as LIGO are sensitive to the mean field quadrature of the GW \cite{guerreiro2022quantum}. Consider the parent mode is populated by some state $ \rho_{a} $, while the nonlinear mode is initially in the vacuum. To linear order in the interaction time $ t $ the amplitude quadrature of mode $ b $ is given by
\begin{eqnarray}
    \langle X_{b}(t) \rangle \approx \frac{\kappa t}{2} \Bigl< X_{a}(0)Y_{a}(0) + Y_{a}(0)X_{a}(0) \Bigr> 
    \label{quadrature}
\end{eqnarray}
where $ X_{a} = a + a^{\dagger} $ and $ Y_{a} = i(a^{\dagger} - a) $ are the amplitude and phase quadratures of the parent mode at $ t = 0 $ and mean values are taken with respect to $ \rho_{a} \otimes \vert 0 \rangle \langle 0 \vert $. Eq. \eqref{quadrature} remains valid to linear order in $ t $ even in the presence of absorption (amplitude damping) for modes $ a $ and $ b $, provided $ b $ is initially in the vacuum; see the Appendix for details. During the early moments of the nonlinear ringdown, we can expect the growth rate of the GW amplitude will be proportional to the amplitude-phase quantum correlations of the parent mode and to leading order in the coupling constant the parent mode will become squeezed \cite{mandel1982squeezing}.

The most general Gaussian single-mode state consists of a displaced-squeezed-thermal state defined in terms of a displacement amplitude $ h = \vert h \vert e^{i\theta} $, a squeezing parameter $ \xi = re^{i\phi} $ and occupation number $ \Bar{n}$ \cite{weedbrook2012gaussian}. In this case, the correlator in the r.h.s of \eqref{quadrature} reads
\begin{align}
    \frac{1}{2}\Bigl< X_{a}Y_{a} + Y_{a}X_{a} \Bigr> = \vert h \vert^{2} \sin \theta \cos \theta \nonumber \\
    + \left(  2\bar{n} + 1  \right) \sinh 2r \sin \phi \cos \phi
\end{align}
The first term is always present for coherent states, while the second term is a correction dependent on $ \bar{n} $ and $ \xi $.
Direct measurements of the nonlinear mode amplitude can thus deviate significantly from the classical prediction, which requires $ \frac{1}{2}\langle XY + YX \rangle = \langle X \rangle \langle Y \rangle  $. Note the correction term is accompanied by enhancement factors given by the thermal number $ \bar{n}$ and the exponential of the squeezing parameter $ e^{\vert \xi \vert} = e^{r} $, the same enhancement factors appearing in the detection of quantum GW states \cite{guerreiro2020gravity, parikh2021a, parikh2021b, guerreiro2022quantum, chawla2023quantum, bak2023quantum}. Recall the thermal nature of a GW state could arise through different mechanisms, for instance via the cosmic GW background or Hawking radiation \cite{hawking1974black} and the mean number of thermal gravitons is related to temperature through $ \bar{n} = 1/(e^{ \hbar \omega_{a} / k_{B} T} - 1) \approx k_{B}T /  \hbar \omega_{a} $ for large $ T $.

Squeezed-coherent states display an intricate second-order correlation function \cite{kanno2019detecting} dependent on the displacement and squeezing parameters: 
they can be sub-Poissonian or super-Poissonian, even for arbitrarily large mean number of gravitons \cite{guerreiro2022quantum}. This opens the possibility of detecting sub-Poissonian statistics of the gravitational field, a macroscopic quantum phenomenon. 

Finally, we emphasize that the dependence of nonlinear modes on quantum properties of parent states goes beyond Eqs. \eqref{OPA1} and \eqref{OPA2}: even if nonlinear ringdowns are corrected by more complex laws, observables will still be sensitive to quantum features of the parent state provided GWs have a corpuscular nature.

\textit{Gravitational entanglement.}-- Frequency doubling can not only act as a detector for the quantum properties of parent GW modes, but also as a source of non-classical gravitational states. Second harmonic nonlinearities generically create entanglement between parent and nonlinear modes starting from a coherent state input \cite{olsen2004continuous, grosse2006harmonic}, particularly in the strong coupling regime \cite{villar2006direct, coelho2009three, dechoum2010semiclassical}.
Numerical simulations of black hole ringdowns show a ratio $ A_{(4,4)}^{(2,2,0)\times(2,2,0)} / (A_{(2,2,0)})^{2} $ of order one \cite{mitman2023nonlinearities}, suggesting that GWs are strongly coupled, $ \vert \kappa \vert / \gamma 	\gtrsim 1 $.
The exact properties of entanglement will depend on the remnant spin parameter, mode coupling, damping rates and angular selection rules between the interacting GW modes, and are therefore difficult to predict analytically. 
We can nevertheless take a phenomenological approach and search for signatures of quantum correlations in ringdown signals. 
We now describe how such signatures can be experimentally searched by monitoring an ensemble of optical states that interact with the wave. 

\begin{figure}[t!]
    \centering
    \includegraphics[width=0.4\textwidth]{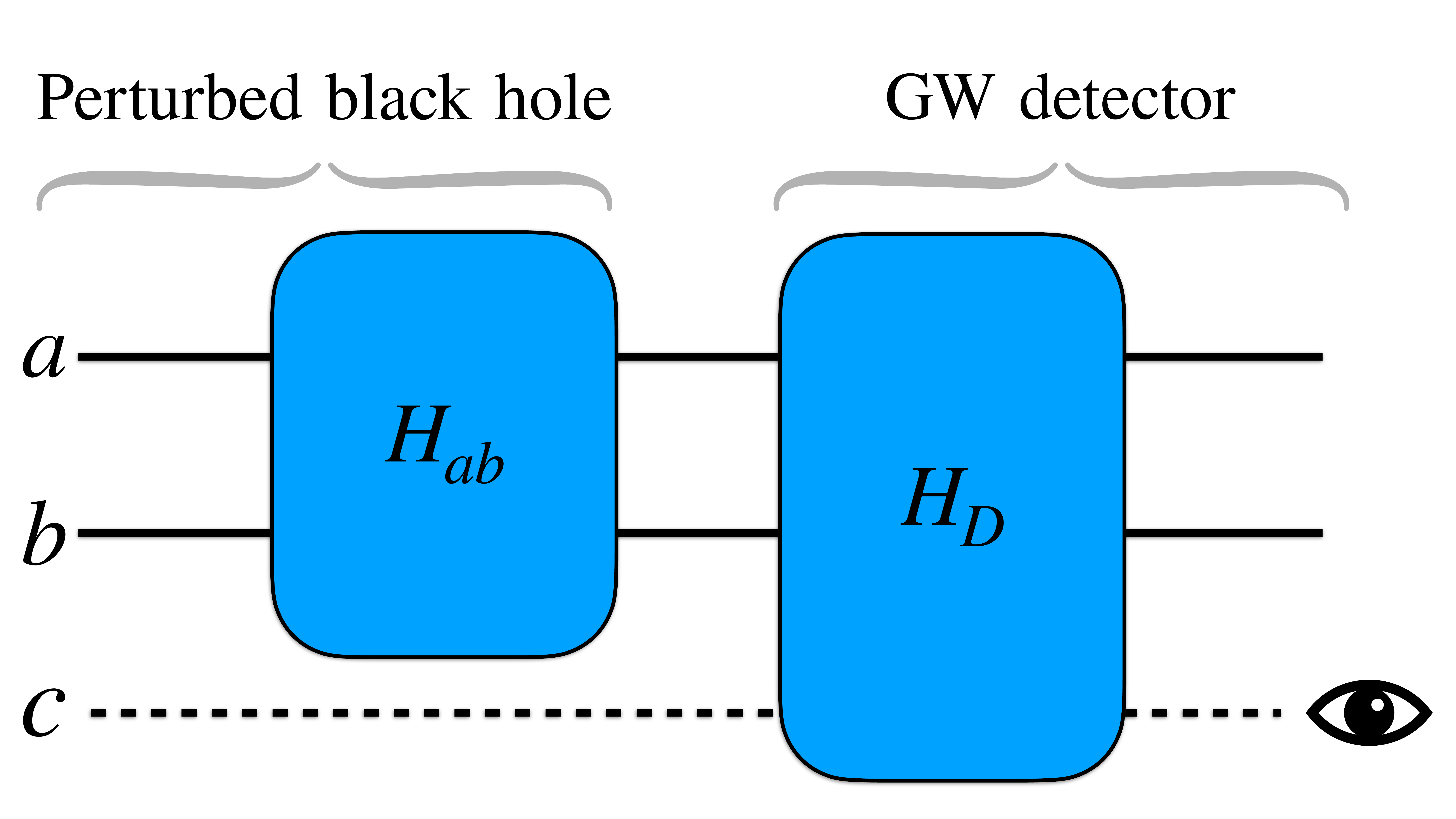}
    \caption{Gravitational wave entanglement witness experiment. Black hole nonlinearities act as a source of entangled GWs by virtue of the interaction Hamiltonian $ H_{ab} $ and amplitude-damping operators $ L_{a} = \sqrt{2\gamma_{a}}a$, $ L_{b} = \sqrt{2\gamma_{b}}b$. Gravitational entanglement can be empirically searched by monitoring an ensemble of optical modes that interact with the waves. Correlations between the GW modes $ a, b$ get imprinted on the reduced density matrix of the optical mode $ c $, according to the interaction $ H_{D} $. 
    }
    \label{fig1}
\end{figure}

Consider a cavity as a simplified model GW interferometer \footnote{This is arguably a simplification, but it captures the essence of a GW detector and notice that it can be formally mapped into a Michelson interferometer such as LIGO; see \cite{demkowicz2015quantum}.}. 
Starting from linearised Einstein's equations in Minkowski background, one can show that the interaction between the cavity and a discrete set of GW modes consists of a direct coupling \cite{pang2018quantum} described by the optomechanical dispersive Hamiltonian in which the GW modes formally assume the role of mechanical oscillators \cite{guerreiro2020gravity, guerreiro2022quantum}. We refer to this as the \textit{optogravitational} interaction. 
For two modes the interaction Hamiltonian reads,
\begin{eqnarray}
    H_{D} = \omega_{c} c^{\dagger}c + \omega_{a}a^{\dagger}a + \omega_{b}b^{\dagger}b - c^{\dagger}c \left(  g_{a}X_{a} + g_{b} X_{b}   \right) \ ,
    \label{two_mode_hamilton}
\end{eqnarray}

\noindent where $ c (c^{\dagger}) $ represents the annihilation (creation) operator for the cavity and $ \omega_{c} $ is the cavity frequency. The optogravitational couplings $ g_{i} $ are given by
\begin{eqnarray}
    g_{i} = \omega_{i}q_{i} = \frac{\omega_{c}}{4} \sqrt{\frac{8\pi G}{\omega_{i}V}} \ ,
\end{eqnarray}
where $ i = a,b $, $ V $ is the quantization volume, $ G $ is Newton's constant and we have introduced the \textit{dimensionless} coupling $ q_{i} $. Note $ \mathrm{max}\lbrace q_{i} \rbrace \approx (\omega_{c} / E_{\rm pl})  $, so GW detectors are weakly coupled optomechanical systems \cite{guerreiro2020gravity}. Together with the free field Hamiltonians for $ a, b $ and $ c$, Eq. \eqref{two_mode_hamilton} can be exponentiated to give the interaction picture time evolution operator \cite{brandao2020entanglement},
\begin{eqnarray}
    U(t) = e^{-iB(t)(c^{\dagger}c)^{2}} e^{q_{a}c^{\dagger}c(\gamma_{a}a^{\dagger} - \gamma_{a}^{*}a)} e^{q_{b}c^{\dagger}c(\gamma_{b}b^{\dagger} - \gamma_{b}^{*}b)}
    \label{unitary}
\end{eqnarray}

\noindent where we have defined,
\begin{eqnarray}
    \gamma_{k} &=& (1 - e^{-i\omega_{k} t}) \label{gammas} \\
    B(t) &=&  \sum_{k = a,b} q_{k}^{2} \left(   \omega_{k} t - \sin \omega_{k} t  \right) \ . \label{Bs}
\end{eqnarray}
Furthermore, one can show that the effect of GW modes populated by the vacuum state yields negligible corrections to Eq. \eqref{unitary}; see Appendix for details.

Consider an ensemble of optical states interacting with GW modes $ a, b $. We are interested in calculating the time development of the reduced optical density matrix, given an initial state of the form
\begin{eqnarray}
    \rho_{cab}(0) = \left(  \sum_{n,m}a_{n}a^{*}_{m} \vert n \rangle \langle m \vert \right) \otimes \sigma_{ab} \ ,
\end{eqnarray}
where $ a_{n} $ are known complex coefficients defining the initial optical density matrix and $ \sigma_{ab} $ is the parent-nonlinear GW joint state. 
A general element of the reduced optical density matrix $\rho_{nm}(t) $ reads
\begin{eqnarray}
    \rho_{nm}(t) = a_{n}a^{*}_{m} e^{-i \delta B(t)} e^{-i\omega_{c}\Delta t} \ \mathcal{C}(t) \ ,
\end{eqnarray}

\noindent where $ n,m \in \mathbb{N} $, $ \delta = (n^{2} - m^{2}) \ , \Delta = (n-m) $ and we have defined the correlation function $ \mathcal{C}(t)$, 
\begin{eqnarray}
    \mathcal{C}(t) = \mathrm{Tr} \left( \sigma_{ab}   e^{\Delta q_{a}(\gamma_{a}^{*}a^{\dagger} - \gamma_{a}a))} e^{\Delta q_{b}(\gamma_{b}^{*}b^{\dagger} - \gamma_{b}b))}  \right) \ .
\end{eqnarray}
Note that $ \rho_{nm}(t) $ consists of a state-independent function of time multiplied by $ \mathcal{C}(t) $, which depends on the initial joint GW state $\sigma_{ab}$. 
Therefore, by measuring the reduced density matrix describing the optical ensemble we can obtain information on the quantum correlations between the GW modes $ a $ and $ b $. 
In particular, to quadratic order in the dimensionless couplings $ q_{i}^{2}, q_{a}q_{b} $, or linear order in $ G $, we find
\begin{widetext}
    \begin{eqnarray}
    \mathcal{C}(t) \approx 1 &-& i q_{a} \Delta \left(   \sin \omega_{a}t \langle X_{a} \rangle + (1 - \cos \omega_{a}t) \langle Y_{a} \rangle  \right)
     \nonumber \\
    &-& \frac{q_{a}^{2}\Delta^{2}}{2} \left(  \sin^{2}\omega_{a}t  \langle X_{a}^{2} \rangle + (1 - \cos \omega_{a}t)^{2} \langle Y_{a}^{2} \rangle + \sin \omega_{a}t (1 - \cos \omega_{a}t) \bigl< X_{a}Y_{a} + Y_{a}X_{a} \bigr>    \right)  \nonumber \\
    & + & \lbrace b-\mathrm{terms} \rbrace \nonumber \\
    &-& q_{a}q_{b}\Delta^{2} \big(  \sin \omega_{a}t \sin \omega_{b}t \langle X_{a} X_{b} \rangle +  \sin \omega_{a}t (1 - \cos \omega_{b}t) \langle X_{a} Y_{b} \rangle  \nonumber \\
    & \ & \ +  \sin \omega_{b}t (1 - \cos \omega_{a}t) \langle Y_{a} X_{b} \rangle +   (1 - \cos \omega_{a}t)(1 - \cos \omega_{b}t) \langle Y_{a} Y_{b} \rangle  \big)
    \label{correlator_linearised}
\end{eqnarray}
\end{widetext}
where expectation values are taken with respect to $ \sigma_{ab} $. Eq. \eqref{correlator_linearised} contains each of the fourteen second-order amplitude-phase correlators for modes $a $ and $ b $, multiplied by independent functions of time. Generically, by collecting the value of $ \mathcal{C}(t) $ (or equivalently $ \rho_{nm}(t) $) at fourteen distinct times we can reconstruct all of the second-order amplitude-phase correlation functions \footnote{When choosing sampling times, care must be taken to avoid the effects of aliasing.}. From these correlators, an entanglement witness can be set up, for example Duan's inseparability criterion:
\begin{eqnarray}
    \mathcal{D} = \mathrm{Var}(X_{a} + X_{b}) + \mathrm{Var}(Y_{a} - Y_{b})  \ ,
\end{eqnarray}
for which $ \mathcal{D} < 4 $ implies entanglement \cite{duan2000inseparability}. 

Strong nonlinear couplings between GWs in perturbed BHs might act as a source of entangled states of the gravitational field. Since GWs interact very weakly with matter, we expect these states will travel the universe with negligible decoherence \cite{zeldovich1983relativistic} until they reach a GW detector and interact via the optogravitational Hamiltonian $ H_{D} $. Quantum correlations of the gravitational modes will then be imprinted upon the ensemble of optical states, for which subsequent measurements can be performed to reveal the gravitational entanglement -- see Fig. \ref{fig1} for a schematic representation.

\textit{Discussion.} -- In a nutshell, we can expect that nonlinear GW signals will depend on quantum properties of parent modes. This can be tested by comparing nonlinear ringdown measurements with classical predictions \cite{loutrel2021second, ripley2021numerical} and expectations from the inspiral phase \footnote{The inspiral phase is well described by classical numerical predictions, which are expected to match the case of coherent states in effective field theory \cite{guerreiro2020gravity}.}.
As in quantum optics, nonlinear
GW production from parent squeezed and thermal states
should be substantially larger than from coherent GWs. 
Other possibly very interesting states are squeezed-thermal, beyond quadratic squeezing \textit{molded} quantum states \cite{parikh2021a} and bunched states exhibiting sub-Poissonian statistics \cite{zou1990photon}.
Because of the simple bunching v.s. anti-bunching argument, the dependence of nonlinear signals on statistical properties of the GWs will also follow for higher-order or multi-mode interactions, as well as for cascaded nonlinear processes that might occur in strong gravity. 
Therefore, it is to be expected that the dependence of nonlinear ringdown modes on quantum effects is robust and goes beyond quadratic effects. This offers a new route to probe the quantum nature of gravity. 

There are plenty of pathways for producing quantum GW states in strong gravity. 
Nonlinear interactions, which are intrinsic to gravity, temper with the graviton number distribution, generically leading to non-classical states. 
Squeezed states for instance can arise from frequency doubling in itself,
both in the parent \cite{mandel1982squeezing, pereira1988generation} and second harmonic waves \cite{paschotta1994bright}.
Moreover, quantum fluctuations could trigger macroscopic revivals in the frequency doubling signals, which would also provide evidence for gravitational quantum fluctuations \cite{olsen2000quantum}. In this case, nonlinearities would act as amplifiers.

Second-order nonlinearities also generate entanglement between modes populated by intense fields. A notable example is the optical parametric oscillator operated above threshold, capable of generating three partite entanglement between intense fields starting form a strong coherent state \cite{villar2006direct, coelho2009three}. The gravitational counterpart of these quantum correlations between parent and nonlinear modes can be revealed through statistical measurements on an ensemble of optical states that interact with the entangled GWs. Second-order nonlinearities can act both as detector and source of quantum GW states.



Nonlinear processes beyond frequency doubling also produce non-classical states. This can be understood on very general grounds as a consequence of conservation laws and can occur even in the presence of dissipation \cite{hillery1985conservation}.
In quantum optics, non-coherent states are commonly generated from parent coherent waves by parametric amplification, a phenomenon that can also take place for gravity. Einstein’s equations imply that the Riemann curvature
tensor satisfies a nonlinear wave equation \footnote{$ R_{\alpha \beta \gamma \delta ; \mu}^{ \ \ \ \ \ \ \ \mu} = 2 R_{\alpha \beta \xi \mu} R_{\delta \ \ \gamma \ }^{ \ \ \xi \ \ \mu} + 2 R_{\alpha \xi \delta \mu} R_{\beta \ \ \gamma \ }^{ \ \ \xi \ \ \mu} - 2 R_{\alpha \xi \gamma \mu} R_{\beta \ \ \delta \ }^{ \ \ \xi \ \ \mu} $}. Splitting the curvature into background and perturbations \cite{brill1964method}, there are two mechanisms for the parametric amplification of GWs. First, the covariant derivative in the wave operator contains connection coefficients which are dependent on the metric perturbation $ h $, thus producing higher-order interaction terms in the GW field schematically of the form $ h\partial h \partial h, h^{2}\partial h \partial h, ... $ \cite{zee2010quantum}. Second, coupling terms between GWs and the dynamical background curvature appear and if the radius of curvature of the background is comparable to the perturbation's wavelength, this can lead to parametric amplification. The latter situation is familiar in inflationary cosmology, where rapid expansion parametrically amplifies the GW vacuum producing a squeezed state \cite{grishchuk1990squeezed, albrecht1994inflation}. It is conceivable that similar mechanisms can arise in highly perturbed BHs.


Another possible path to non-coherent GW states is absorption-induced mode excitation \cite{sberna2022nonlinear}: a fast-decaying parent mode is absorbed by the BH thus changing its mass. This modifies the background spacetime, causing the remaining parent mode to be projected onto a new QNM spectrum. These sudden changes in frequency lead to squeezing -- in fact, this is in essence analogous to the background curvature coupling that generates squeezed states in inflationary models \cite{allen1996stochastic}.

We can think of perturbed BHs as a temporary storage of GWs \cite{goebel1972}.
The imaginary part of the QNM frequency dictates the quality factor $ Q $ of the resonator. Quantum features such as the degree of squeezing, graviton statistics and entanglement of generated states will depend on $ Q $. Absorption by the BH is analogous to loss in a cavity, hence a form of decoherence. 
There exists modes of Kerr BHs for which the damping vanishes asymptotically as the spin approaches the extremal value \cite{yang2013branching}. Thus, high spin will likely stand out in the search for quantum effects in nonlinear GWs. Finally, we cannot resist sketching an \textit{optomechanical} model for the generation of nonlinear GWs: BHs can be thought of as an oscillating droplet, with GWs populating whispering-gallery modes. Perturbations, including those due to coexisting QNMs, effectively change the whispering gallery's effective path length, which leads to optomechanical squeezing -- see Fig.~1 of \cite{childress2017cavity} for a schematic illustration.

Nonlinear ringdowns have not yet been observed, but the next generation of detectors will likely change that \cite{maggiore2020science, evans2021horizon}. Classical general relativity might not be sufficient to predict the properties of those ringdowns. BH spectroscopy would then give experimental evidence on the quantum nature of strong gravity, similarly to how optical spectroscopy led to quantum mechanics.
The strong nonlinear character of gravity might also produce higher order or multi-mode couplings and cascaded nonlinear processes capable of preparing complex non-classical states, opening an experimental window into the quantum nature of space-time.

\acknowledgments{The author acknowledges Guilherme Tempor\~ao, Antonio Zelaquett Khoury, George Svetlichny, Carlos Tomei, Luca Abrah\~ao, Igor Califrer, Francesco Coradeschi and Antonia Micol Frassino for conversations. This work was supported in part by the Coordenac\~ao de Aperfei\c{c}oamento de Pessoal de N\'ivel Superior - Brasil (CAPES) - Finance Code 001, Conselho Nacional de Desenvolvimento Cient\'ifico e Tecnol\'ogico (CNPq), Funda\c{c}\~ao de Amparo \`a Pesquisa do Estado do Rio de Janeiro (FAPERJ Scholarship No. E-26/202.830/2019) and Funda\c{c}\~ao de Amparo \`a Pesquisa do Estado de São Paulo (FAPESP processo 2021/06736-5).}

\bibliography{main}

\appendix

\onecolumngrid


\section{Energy conservation and frequency doubling}

Let $ a \ (a^{\dagger}) $, $ b \ (b^{\dagger}) $ denote annihilation (creation) operators for two Bosonic modes, satisfying the canonical commutation relations,
\begin{align}
    [a,a^{\dagger}] = [b^{\dagger}, b] = \mathds{1} 
\end{align}
\begin{align}
     [a, b] = [a^{\dagger}, b] = [a, b^{\dagger}] = [a^{\dagger}, b^{\dagger}] = 0
\end{align}
Define number operators $ N_{a} = a^{\dagger}a , N_{b} = b^{\dagger}b $. We want to find the most general Hermitian operator composed of at most third-order products of creation and annihilation operators that commutes with 
\begin{eqnarray}
    J = N_{a} + 2N_{b}
\end{eqnarray}
For convenience we list all the commutators with third-order operators:
\begin{table}[h!]
\begin{tabular}{lllll}
                                    &                                                                                                       &                                                                                  &                                                                                                        &                                                                                  \\
$[a,J] = a$                         & $[aa,J] = 2aa$                                                                                        & $ [aaa,J] = 3aaa $                                                               & $ [aab,J] = 4aab $                                                                                     & $ [abb,J] = 5abb $                                                               \\
$[a^{\dagger},J] = - a^{\dagger}$   & $ [bb,J] = 4bb $                                                                                      & $ [aaa^{\dagger},J] = aaa^{\dagger} $                                            & $ [aab^{\dagger},J] = 0 $                                                                              & $ [abb^{\dagger},J] = abb^{\dagger} $                                            \\
$ [b,J] = 2 b $                     & $ [ab,J] = 3ab $                                                                                      & $ [aa^{\dagger}a,J] = aa^{\dagger}a $                                            & $ [a^{\dagger}a^{\dagger}b,J] = 0 $                                                                    & $ [ab^{\dagger}b,J] = ab^{\dagger}b $                                            \\
$ [b^{\dagger},J] = -2b^{\dagger} $ & $ [a^{\dagger}b,J] = a^{\dagger}b $                                                                   & $ [a^{\dagger}aa,J] = a^{\dagger}aa $                                            & $ [aa^{\dagger}b,J] = 2aa^{\dagger}b = 2a^{\dagger}ab+2b $                                             & $ [a^{\dagger}bb,J] = 3a^{\dagger}bb $                                            \\
                                    & $ [ab^{\dagger},J] = -ab^{\dagger} $                                                                  & $ [aa^{\dagger}a^{\dagger}, J] = -  aa^{\dagger}a^{\dagger} $                    & $ [a^{\dagger}ab,J] = 2a^{\dagger}ab $                                                                 & $ [ab^{\dagger}b^{\dagger},J] = - 3ab^{\dagger}b^{\dagger} $                     \\
                                    & $ [a^{\dagger}b^{\dagger},J] = -3a^{\dagger}b^{\dagger} $                                             & $ [a^{\dagger}aa^{\dagger},J] = - a^{\dagger}aa^{\dagger} $                      & $ [aa^{\dagger}b^{\dagger},J] = -2aa^{\dagger}b^{\dagger} = -2a^{\dagger}ab^{\dagger} - 2b^{\dagger} $ & $ [a^{\dagger}bb^{\dagger},J] = - a^{\dagger}bb^{\dagger} $                      \\
                                    & $ [a^{\dagger}a^{\dagger},J] = - 2 a^{\dagger}a^{\dagger} $ & $ [a^{\dagger}a^{\dagger}a,J] = - a^{\dagger}a^{\dagger}a $                      & $ [a^{\dagger}ab^{\dagger},J] = -2a^{\dagger}ab^{\dagger} $                                            & $ [a^{\dagger}b^{\dagger}b,J] = - a^{\dagger}b^{\dagger}b $                      \\ 
                                    & $ [b^{\dagger}b^{\dagger},J] = - 4 b^{\dagger}b^{\dagger} $                                           & $ [a^{\dagger}a^{\dagger}a^{\dagger},J] = - 3a^{\dagger}a^{\dagger}a^{\dagger} $ & $ [a^{\dagger}a^{\dagger}b^{\dagger},J] = -4 a^{\dagger}a^{\dagger}b^{\dagger} $                       & $ [a^{\dagger}b^{\dagger}b^{\dagger},J] = -5 a^{\dagger}b^{\dagger}b^{\dagger} $
\end{tabular}
\end{table}

Note that commutators with triple $ b$'s will have the same values as those with triple $ a$'s with an extra factor of 2.
The commutator defines a linear transformation $ T(\  \_ \ ) = [\ \_ \ ,J] $ from Hermitian to anti-Hermitian operators. By inspection, we can see the kernel of $ T $ is given by $ H_{ab} $ as defined in the main text.  
Energy conservation is sufficient to single out the frequency doubling Hamiltonian.


\section{Evolution of quadrature operators}

\subsection{Equations of motion}

Consider the Markovian open quantum system evolution for an operator $\mathcal{O} $, 
\begin{equation}
    \dot{\mathcal{O}} = i [H,\mathcal{O}] + \sum_{\nu} \left(  L_{\nu}^{\dagger} \mathcal{O} L_{\nu} - \frac{1}{2} L_{\nu}^{\dagger} L_{\nu}\mathcal{O} - \frac{1}{2} \mathcal{O} L_{\nu}^{\dagger} L_{\nu} \right) \ .
    \label{lindblad}
\end{equation}
We are interested in studying the evolution of the creation and annihilation operators of two Bosonic modes $ a, b $ evolving according to 
\begin{equation}
    H = H_{a} + H_{b} + H_{ab} ,
\end{equation}
where 
\begin{equation}
    H_{a} = \omega a^{\dagger}a \ , H_{b} = 2\omega b^{\dagger} b \ , H_{ab} = \kappa \left(  b^{\dagger}a^{2} + b a^{\dagger 2}  \right) 
\end{equation}
and 
\begin{equation}
    L_{a} = \sqrt{2\gamma_{a}} a \ , L_{b} = \sqrt{2\gamma_{b}} b \ .
\end{equation}

Define the operators $ A $ and $ B $ according to,
\begin{equation}
    \alpha = ae^{i\omega t} \ , \beta = b e^{2i\omega t}
\end{equation}
Observe $ \alpha $ and $ \beta $ have the same commutation relations as $ a, b$. 
Eq. \eqref{lindblad} for modes $ a $ and $ b $ reads,
\begin{eqnarray}
    \dot{\alpha} &=& -2i\kappa \beta \alpha^{\dagger} - \gamma_{a}\alpha \\
    \dot{\beta} &=& -i\kappa \alpha^{2} - \gamma_{b}\beta
\end{eqnarray}
We can also find the corresponding second order Eqs. in the rotating frame:
\begin{eqnarray}
    \ddot{\alpha} &=& -2\kappa^{2} \left( N_{a} - 2N_{b} \right) \alpha - 2i\kappa \left( \gamma_{b} - \gamma_{a}   \right)\beta \alpha^{\dagger} - \gamma_{a}\dot{\alpha} \label{daughter} \\ 
    \ddot{\beta} &=& -4\kappa^{2}\left(   N_{a} + \frac{1}{2} \right) \beta + 4i\kappa \gamma_{a}\alpha^{2} - \gamma_{b}\dot{\beta} \label{parent}
\end{eqnarray}
Note that when $ \gamma_{a} = \gamma_{b} = \gamma $ we can cast Eq. \eqref{daughter} in the form of a parametrically driven oscillator,
\begin{eqnarray}
    \ddot{q} + \gamma \dot{q} + \omega^{2}\left(   1 + f(t)\right)q = 0 \label{parametric}
\end{eqnarray}
Eq. \eqref{parametric} has been discussed in \cite{yang2015turbulent} in the context of nonlinear mode interactions of perturbations of a Kerr black hole originating from a curvature background-perturbations coupling.

\subsection{Quadrature}

We are interested in the behavior of $ \beta(t) $ for small interaction times. To linear order in $ t $ \cite{kozierowski1977quantum, mandel1982squeezing},
\begin{eqnarray}
    \beta(t) &=& \beta(0) + t \dot{\beta}(0) + \frac{t^{2}}{2}\ddot{\beta}(0) + ... \\
    &=& \beta(0) + t \left( - i\kappa \alpha^{2}(0) - \gamma_{b}\beta(0) \right) + \mathcal{O}(t^{2}) 
\end{eqnarray}
Consider the initial state for modes $ a, b $ is $ \vert \Psi\rangle_{a} \vert 0 \rangle_{b} $. Then, $ \langle \beta(0) \rangle = \langle \beta^{\dagger}(0) \rangle = 0  $, and we can write,
\begin{eqnarray}
    \langle \beta(t) \rangle \approx -i\kappa t \langle \alpha^{2}(0) \rangle
\end{eqnarray}
Define the field quadratures
\begin{eqnarray}
    X_{z}(t) = z e^{i\omega_{z}t} + z^{\dagger} e^{-i\omega_{z}t} \ , \ 
    Y_{z}(t) = i\left(  z^{\dagger} e^{-i\omega_{z}t} - z e^{i\omega_{z}t} \right)
    \label{quadratures_times}
\end{eqnarray}
where $ z = a,b $. To first order in $ t $ we have,
\begin{eqnarray}
    \langle X_{a}(t) \rangle \approx \frac{\kappa t}{2} \Bigl< X_{a}(0)Y_{a}(0) + Y_{a}(0)X_{a}(0) \Bigr> 
\end{eqnarray}

A Gaussian state is completely determined by the mean values of the field quadratures and their covariance matrix $ \mathds{V} $. For a single mode state with quadrature operators $ X,Y $ we have $ \mathds{V}_{XY} = \frac{1}{2}  \Bigl< XY + YX\Bigr> - \langle X \rangle \langle Y \rangle $. The most general Gaussian state corresponds to a displaced, rotated, squeezed state, and has covariance matrix \cite{weedbrook2012gaussian},
\begin{eqnarray}
    \mathds{V} = (2\bar{n}+ 1) \mathbf{R}(\phi) \mathbf{S}(2r) \mathbf{R}(\phi)^{T}
    \label{covariance}
\end{eqnarray}
where $ \mathbf{R}(\phi) $ is a $ 2 \times 2 $ rotation matrix, $ \mathbf{S}(2r) = \mathrm{diag}(e^{2r}, e^{-2r}) $ is the squeezing matrix with parameter $ r $, and $ \bar{n}$ is the thermal state mean occupation number. From the definition of the covariance matrix and Eq. \eqref{covariance} we have,
\begin{eqnarray}
    \frac{1}{2}  \Bigl< XY + YX\Bigr> = (2\bar{n}+ 1)\sinh r \sin \phi \cos \phi + \langle X \rangle \langle Y \rangle 
\end{eqnarray}
where $ \langle X \rangle = \vert h \vert \cos \theta $ and $ \langle Y \rangle = \vert h \vert \sin \theta  $ are the mean quadratures of a coherent state with amplitude $ he^{i\theta} $.
When $ \bar{n} = 0 $ we get the most general pure Gaussian state, corresponding to a rotated squeezed-coherent state,
\begin{eqnarray}
    \vert \Psi \rangle = D(h)R(\phi)S(r)\vert 0 \rangle
\end{eqnarray}
where $ D, R, S $ denote the displacement, phase rotation and squeezing operators, respectively.

\section{Interaction between GW modes and cavity}

We briefly review the interaction Hamiltonian between a GW and a model GW detector and detail some calculations used in the main text. We follow the treatment outlined in \cite{pang2018quantum, guerreiro2020gravity, guerreiro2022quantum}.

\subsection{Hamiltonian}

Consider an optical cavity as a model for the GW detector. 
Starting from linearised Einstein's equations in Minkowski background, one can show that the interaction consists of a direct coupling between the optical modes and the GW perturbations \cite{pang2018quantum}.
The total Hamiltonian for the GW + optical modes is given by
\begin{eqnarray}
    H = H_{0} + H_{D}
\end{eqnarray}
where $ H_{0} $ are the free field Hamiltonians (to be discussed later) and $ H_{D} $ describes the interaction between the GW modes and the detector. 
For a $ + $ polarized GW propagating perpendicular to the cavity axis the interaction is given by 
\begin{eqnarray}
H_{D} = - \dfrac{\omega_{c}}{4} c^{\dagger} c \int \dfrac{d \bm{k}}{\sqrt{(2\pi)^{3}}} \left(   \sqrt{\dfrac{8\pi G}{k}} \mathfrak{b}_{\bm{k}} + h.c.  \right) \ ,
\label{H_int_canonical}
\end{eqnarray}  
where $ k = \vert \bm{k} \vert = \omega_{k} $ is the GW frequency for the mode $ \bm{k} $, the operator $  \mathfrak{b}_{k}^{\lambda} $ ($ \mathfrak{b}_{\bm{k}}^{\lambda \dagger} $) is the canonical graviton annihilation (creation) operator and $ \omega_{c} $ is the cavity frequency with photon annihilation (creation) operator $ c \ (c^{\dagger}) $.
To obtain a discrete version of \eqref{H_int_canonical} we introduce a quantization volume $ V $ and note that $ \left[ \sqrt{8\pi G / k} \right] = L^{3/2} $, where $ L$ denotes dimension of length. Moreover, $ \left[ d \bm{k} \right] = L^{-3} $. The graviton annihilation and creation operators then have dimension $ \left[ \mathfrak{b}_{\bm{k}} \right] = L^{3/2} $. Define $ \mathfrak{b}_{\bm{k}} = \sqrt{V}  \bm{b}_{\bm{k}} $, where $  \bm{b}_{\bm{k}} $ is dimensionless.  The discrete limit $ d \bm{k} \rightarrow 1 / V $, $ (2\pi)^{-3/2}\int \rightarrow \sum $ leads to, 
\begin{eqnarray}\label{eq:H_I}
H_{D} = - \dfrac{\omega_{c}}{4} c^{\dagger} c \sum_{\bm{k}} \left(   \sqrt{\dfrac{8\pi G}{kV}} \bm{b}_{\bm{k}} + h.c. \right).
\end{eqnarray}
Defining the single graviton strain for mode $ \bm{k} $ as $ f_{\bm{k}} =  \sqrt{8\pi G/(kV)} $, then the coupling constant of a GW mode $ \bm{k} $ with the cavity electromagnetic field is $ g_{\bm{k}} = \omega_{c}  f_{\bm{k}} / 4 $, which has dimension of frequency.
It is also convenient to introduce the dimensionless coupling $ q_{\bm{k}} =  g_{\bm{k}} / \omega_{k} $.

The discretized free-field Hamiltonian reads
\begin{eqnarray}\label{eq:H_0}
H_{0} = \omega_{c} c^{\dagger} c + \sum_{\bm{k}} \omega_{k} \bm{b}_{\bm{k}}^{\dagger} \bm{b}_{\bm{k}} \,.
\end{eqnarray}

In the interaction picture, the time evolution is given by 
\begin{eqnarray}
U(t) = e^{-i H t} = \prod_{\bm{k}} U_{\bm{k}}(t)\,,
\label{total_U}
\end{eqnarray}
where,
\begin{eqnarray}
U_{\bm{k}} (t)  = e^{iB_{k}(t) (c^{\dagger}c)^{2}} e^{q_{\bm{k}} c^{\dagger} c  (\gamma_{k} \bm{b}^{\dagger}_{k} -\gamma_{k}^{*} \bm{b}_{k})} \,,
\label{Uk}
\end{eqnarray}
with 
\begin{eqnarray}
    \gamma_{k} &=& (1 - e^{-i\omega_{k} t}) \label{gammas} \\
    B_{k}(t) &=& q_{k}^{2} \left(   \omega_{k} t - \sin \omega_{k} t  \right) \label{Bs}
\end{eqnarray}
and GW states evolve according to \cite{brandao2020entanglement},
\begin{eqnarray}
\vert \Psi(t) \rangle = \prod_{\bm{k}} e^{-i \bm{b}_{\bm{k}}^{\dagger} \bm{b}_{\bm{k}} \omega_{\bm{k}} t} \vert \Psi \rangle \,.
\end{eqnarray}

\subsection{Multimode vacuum corrections and few-mode Hamiltonian}

The evolution operator \eqref{total_U} contains contributions from all modes of the gravitational field, most of which are in the vacuum state. Following \cite{guerreiro2020gravity}, we now show that the contribution due to empty modes can be safely neglected, as expected.
The evolution of the cavity annihilation operator is given by,
\begin{eqnarray}
c(t) &=& \left( \prod_{k} U_{k}(t)   \right)^{\dagger} c \left( \prod_{k} U_{k}(t)   \right) \nonumber \\
&=& \prod_{k} e^{iB_{k}(t) c^{\dagger}c} e^{i B_{k}(t)/2} e^{q_{k}c^{\dagger}c\left( \gamma_{k}\bm{b}^{\dagger} - \gamma_{k}^{*} \bm{b}_{k}   \right)} \ c  \nonumber \\
&=& \prod_{k} e^{iB_{k}(t) c^{\dagger}c} e^{i B_{k}(t)/2} D(q_{k}\gamma_{k}) \ c
\label{annihilation_op}
\end{eqnarray}
where $ D(q_{k}\gamma_{k}) $ is the GW displacement operator. Consider now the GW field to be in the vacuum state
\begin{eqnarray}
\vert 0 \rangle = \prod_{k} \vert 0_{k} \rangle 
\end{eqnarray}
where $ \vert 0_{k} \rangle  $ denotes the vacuum state in mode $ k $. The annihilation operator of an optical mode interacting with such gravitational vacuum can then be written as 
\begin{eqnarray}
c(t) = e^{i \mathcal{F}(t) c^{\dagger} c} e^{i\mathcal{F}(t)/2} \mathcal{G}(t) \ c
\end{eqnarray}
where 
\begin{eqnarray}
\mathcal{F}(t) = \sum_{k} B_{k}(t)
\end{eqnarray}
and 
\begin{eqnarray}
 \mathcal{G}(t) = \prod_{k} \langle 0_{k} \vert D(q_{k}\gamma_{k}) \vert 0_{k} \rangle = \exp \left[ -\dfrac{1}{2} \sum_{k} q_{k}^{2}\vert \gamma_{k}\vert^{2} \right]
\end{eqnarray}
On physical grounds, these expressions must be cut-off at a maximum and minimum frequencies $ \omega_{k} = \vert k \vert $ defining the interval over which the detector is sensitive \cite{parikh2021b}; consider for the purpose of illustration these infrared and ultraviolet cut-offs as the Hubble and Planck energies, $ E_{\rm IR} $ and $ E_{\rm pl}$, respectively. Recovering the continuous limit gives us, up to numerical factors of order one,
\begin{eqnarray}
\mathcal{F}(t) = \int \dfrac{d^{3} \bm{k}}{\sqrt{(2\pi)^{3}}} 2\omega_{c}^{2} \left( \dfrac{8\pi G}{k^{3}} \right) \left( \omega_{k} t - \sin \omega_{k} t  \right)
 \approx  \left( \dfrac{\omega_{c}}{E_{\rm pl}}   \right)^{2} \left( \int_{E_{\rm IR}}^{E_{\rm pl}} dk   \right) t \approx \left( \dfrac{\omega_{c}}{E_{\rm pl}}   \right) \omega t \label{phase_noise}
\end{eqnarray}
where we have used $ E_{\rm pl} \gg E_{\rm IR} $, considered large times $ t \gg E_{\rm IR}^{-1} $ and neglected the bounded term $ \sin \omega_{k} t  $.
Corrections to the optical annihilation operator due to the phase $ \mathcal{F}(t) $ are then on the order of $ (\omega_{c} / E_{\rm pl}) $ and only become relevant for optical frequencies close to the Planck energy.
Similarly,
\begin{eqnarray}
\mathcal{G}(t) = \exp\left(  -\dfrac{1}{2} \int  \dfrac{d^{3} \bm{k}}{\sqrt{(2\pi)^{3}}} \omega_{c}^{2} \left( \dfrac{8\pi G}{k^{3}} \right) \vert \gamma_{k} \vert^{2} \right)
 \approx  \exp \left(  - 2 \left( \dfrac{\omega_{c}}{E_{\rm pl}}   \right)^{2}  \int_{E_{\rm IR}}^{E_{\rm pl}} \dfrac{dk}{k}  \right)
= \exp \left(  - 2\left( \dfrac{\omega_{c}}{E_{\rm pl}}   \right)^{2} \ln \left( \dfrac{E_{\rm pl}}{E_{\rm IR}} \right) \right)
\end{eqnarray}
where we have considered the worst-case approximation $ \vert \gamma_{k} \vert^{2} = 2 \left( 1 - \cos \Omega_{k} t \right) \sim 4 $. Notice that the ratio of ultraviolet to infrared cut-offs is $ E_{\rm pl} / E_{\rm IR} \approx 10^{62} $, giving a correction to $ c(t) $ approximately proportional to $ (\omega_{c} / E_{\rm pl})^{2} $.

Instead of the vacuum, we could consider all modes to be populated by a thermal state at about \SI{1}{K}, the expected temperature for the cosmic GW background \cite{allen1996stochastic}. This would not alter $ \mathcal{F}(t) $, which is state-independent, so the estimates in \eqref{phase_noise} remain. The $ \mathcal{G}(t) $ term, however, would acquire a correction factor at most $ e^{-q_{\rm pl}^{2} (2 \bar{n} + 1)} $, with $ \bar{n} = 1 / (e^{\hbar \Omega_{k} / k_{B}T} - 1) $ and $ q_{\rm pl}
$ the GW coupling strength at the Planck frequency \cite{guerreiro2020gravity}. For a GW mode of \SI{10}{Hz}, peak of the expected cosmic GW background spectrum, this correction factor amounts to $ \approx e^{-10^{-47}} $. 

All in all, these estimates show that the effect of modes which are not populated by states with a large mean number of gravitons upon optical observables is negligible. In other words, decoherence due to the gravitational vacuum, or even due to the thermal background of GWs is very weak, which is consistent with previous results \cite{dyson2014graviton,  blencowe2013effective}. 

\subsection{Few mode Hamiltonian}

From now on, we will take a \textit{few mode} approximation and only consider the terms in the interaction Hamiltonian for which the GW mode is non-empty. For our parent and nonlinear ringdown modes we will consider 
\begin{eqnarray}
    H_{D} = \omega_{c} c^{\dagger}c + \omega_{a}a^{\dagger}a + \omega_{b}b^{\dagger}b - c^{\dagger}c \left(  g_{a}X_{a} + g_{b} X_{b}   \right)
\end{eqnarray}
where the GW parent and nonlinear modes are denoted as $ a $ and $ b $, with frequencies $ \omega_{a} $ and $ \omega_{b} = 2\omega_{a} $, respectively. We have defined amplitude quadrature operators for modes $ a $ and $ b $,
\begin{eqnarray}
    X_{a} = a + a^{\dagger} \ , \ X_{b} = b + b^{\dagger} 
\end{eqnarray}
Recall also the definition of phase operators as,
\begin{eqnarray}
    Y_{a} = i(a^{\dagger} - a) \ , \ Y_{b} = i(b^{\dagger} - b)
\end{eqnarray}
in accordance do Eq. \eqref{quadratures_times}.
Note the two modes couple differently to the GW detector; we have,
\begin{eqnarray}
    \frac{q_{a}}{q_{b}} = \left(\frac{\omega_{b}}{\omega_{a}}\right)^{3/2} = 2\sqrt{2}
\end{eqnarray}
The total unitary evolution in the interaction picture reads,
\begin{eqnarray}
    U(t) = e^{-iB(t)(c^{\dagger}c)^{2}} e^{q_{a}c^{\dagger}c(\gamma_{a}a^{\dagger} - \gamma_{a}^{*}a)} e^{q_{b}c^{\dagger}c(\gamma_{b}b^{\dagger} - \gamma_{b}^{*}b)}
\end{eqnarray}
with $ B(t) = \sum_{k = a,b} B_{k}(t) $, and $ \gamma_{k}, B_{k}(t) $ are defined in Eqs. \eqref{gammas} and \eqref{Bs} above. 

\subsection{Optical density matrix}

Consider an initial optical + GW state of the form 
\begin{eqnarray}
    \vert \psi_{0} \rangle = \vert \Phi \rangle \vert \Psi \rangle
\end{eqnarray}
where $ \vert \Psi \rangle $ is the GW state and $ \vert \Phi \rangle $ is a general optical pure state given by
\begin{eqnarray}
    \vert \Phi \rangle = \sum_{n} a_{n} \vert n \rangle
\end{eqnarray}
We are interested in calculating how the elements of the reduced optical optical density matrix depend on the initial GW state. From measurements of the optical density matrix, we can then obtain information on the parent and nonlinear GW modes, and in particular some of their correlation functions. 

The general optical density matrix element $\rho_{nm}(t) $ reads
\begin{eqnarray}
    \rho_{nm}(t) = a_{n}a^{*}_{m} e^{-i \delta B(t)} e^{-i\omega_{c}\Delta t} \ \mathcal{C}(t)
\end{eqnarray}
where $ \delta = (n^{2} - m^{2}) \ , \Delta = (n-m) $ and we have defined the \textit{correlation function} $ \mathcal{C}(t)$, 
\begin{eqnarray}
    \mathcal{C}(t) &=& \langle \Psi(t) \vert e^{\Delta q_{a}(\gamma_{a}a^{\dagger} - \gamma_{a}^{*}a))} e^{\Delta q_{b}(\gamma_{b}b^{\dagger} - \gamma_{b}^{*}b))} \vert \Psi(t) \rangle = \\
    & = & \langle \Psi \vert e^{\Delta q_{a}(\gamma_{a}^{*}a^{\dagger} - \gamma_{a}a))} e^{\Delta q_{b}(\gamma_{b}^{*}b^{\dagger} - \gamma_{b}b))} \vert \Psi \rangle
    \label{correlator}
\end{eqnarray}

Notice that the dependence of $ \rho_{nm}(t) $ on the initial GW state $ \vert \Psi \rangle $ is entirely contained in $ \mathcal{C}(t)$, with the pre-factor being a known time-dependent function which is the same for any GW state. Therefore, if we know the initial optical state $ \vert \Phi \rangle $ (and its decomposition in terms of the number basis) and the GW modes frequencies, by measuring the optical density matrix we can obtain information on the parent and nonlinear GW correlations. From now on we will treat $ \rho_{nm}(t) $ and $ \mathcal{C}(t)$ interchangeably, in the sense that knowing one implies knowledge of the other. Measuring the optical density matrix therefore allows for measurements of the GWs' correlations. Note that we can generalize \eqref{correlator} to mixed states. For a general initial mixed GW state $ \sigma  $ (unentangled with the optical mode) we have,
\begin{eqnarray}
    \mathcal{C}(t) = \mathrm{Tr} \left( \sigma   e^{\Delta q_{a}(\gamma_{a}^{*}a^{\dagger} - \gamma_{a}a))} e^{\Delta q_{b}(\gamma_{b}^{*}b^{\dagger} - \gamma_{b}b))}  \right)
\end{eqnarray}

Expanding $ \mathcal{C}(t) $ to quadratic order in the couplings $ q_{a}, q_{b} $ we get,
\begin{eqnarray}
    \mathcal{C}(t) \approx 1 &-& i q_{a} \Delta \left(   \sin \omega_{a}t \langle X_{a} \rangle + (1 - \cos \omega_{a}t) \langle Y_{a} \rangle  \right)
     \nonumber \\
    &-& \frac{q_{a}^{2}\Delta^{2}}{2} \left(  \sin^{2}\omega_{a}t  \langle X_{a}^{2} \rangle + (1 - \cos \omega_{a}t)^{2} \langle Y_{a}^{2} \rangle + \sin \omega_{a}t (1 - \cos \omega_{a}t) \bigl< X_{a}Y_{a} + Y_{a}X_{a} \bigr>    \right)  \nonumber \\
    &-& i q_{b} \Delta \left(   \sin \omega_{b}t \langle X_{b} \rangle + (1 - \cos \omega_{b}t) \langle Y_{b} \rangle  \right)
     \nonumber \\
    &-& \frac{q_{b}^{2}\Delta^{2}}{2} \left(  \sin^{2}\omega_{b}t  \langle X_{b}^{2} \rangle + (1 - \cos \omega_{b}t)^{2} \langle Y_{b}^{2} \rangle + \sin \omega_{b}t (1 - \cos \omega_{b}t) \bigl< X_{b}Y_{b} + Y_{b}X_{b} \bigr>    \right)  \nonumber \\
    &-& q_{a}q_{b}\Delta^{2} \big(  \sin \omega_{a}t \sin \omega_{b}t \langle X_{a} X_{b} \rangle +  \sin \omega_{a}t (1 - \cos \omega_{b}t) \langle X_{a} Y_{b} \rangle  \nonumber \\
    & \ & \ +  \sin \omega_{b}t (1 - \cos \omega_{a}t) \langle Y_{a} X_{b} \rangle +   (1 - \cos \omega_{a}t)(1 - \cos \omega_{b}t) \langle Y_{a} Y_{b} \rangle  \big)
\end{eqnarray}
We see this contains terms corresponding to each one of the fourteen second-order amplitude and phase correlators for modes $a $ and $ b $ multiplied by independent functions of time. By collecting the value to $ \mathcal{C}(t) $ (or equivalently $ \rho_{nm}(t) $) at fourteen distinct times, taking care to avoid aliasing of the coefficient functions, we can obtain the correlators necessary to verify the Duan criteria for the parent and nonlinear GW modes.

\end{document}